\documentclass{article}
\usepackage[utf8]{inputenc}
\usepackage{authblk}
\usepackage{footnote}
\usepackage{amsmath}
\usepackage{graphicx}
\usepackage{url} 

\title{Entangled in Spacetime}
\author[1]{Mohammad Rasoolinejad\thanks{Email: Rasoolinejad@u.northwestern.edu}}
\date{September 1, 2024}

\affil[1]{Ph.D. Graduate, Robert R. McCormick School of Engineering and Applied Science, Northwestern University}

\begin{document}
\maketitle

\begin{abstract}
This paper presents an observational analysis of the Delayed-Choice Quantum Eraser experiment through the framework of quantum mechanics. The Delayed-Choice Quantum Eraser, a variation of the classic double-slit experiment, demonstrates the intricate relationship between quantum measurement, wave-particle duality, and the temporal ordering of observations. By utilizing the principles of quantum superposition, entanglement, and the non-local collapse of the wave function, we seek to rationalize the counterintuitive outcomes observed in the experiment. Specifically, we explore how the act of measurement retroactively influences the observed behavior of particles, depending on whether or not the which-path information is available. Our analysis underscores the significance of the quantum mechanical concept of wave function collapse across spacetime, providing a deeper understanding of how quantum mechanics reconciles the delayed-choice paradox.
\end{abstract}

\section{History of Atomic Theory}

The development of atomic theory spans centuries, originating from ancient philosophical conjectures to rigorous modern scientific principles. This evolution has significantly shaped our understanding of the structure and behavior of matter, pivotal to advancements in various branches of science including physics, chemistry, and material sciences. The notion of atoms as indivisible units of matter was first postulated in ancient times, with the term `atom' derived from the Greek word \textit{atomos}, meaning uncuttable. These early ideas, while lacking empirical support, set the groundwork for a more scientific approach that would emerge millennia later. In the 19th century, John Dalton reintroduced the concept of atoms, this time as part of a scientific theory rather than philosophical speculation. Dalton's atomic theory proposed that each element is composed of unique types of atoms and that these atoms combine in simple whole-number ratios to form compounds. His ideas were the first to provide a quantitative framework for chemistry through the law of multiple proportions and the law of conservation of mass, established by Antoine Lavoisier. Lavoisier's experiments refuted the phlogiston theory of combustion and demonstrated the decomposability of water into hydrogen and oxygen, suggesting that these substances were themselves composed of atoms. Joseph Proust expanded on this foundation with the law of definite proportions, which posited that chemical compounds are formed from elements in fixed mass ratios. The combined efforts of these scientists laid the essential groundwork for modern chemistry and atomic physics.

The discovery of subatomic particles began with J.J. Thomson's identification of the electron in 1897. His experiments with cathode rays revealed that atoms were not indivisible, but contained smaller, negatively charged electrons. This finding led to the plum pudding model, which envisioned electrons embedded within a positively charged substrate. However, this model was short-lived, overturned by Ernest Rutherford's gold foil experiment in 1911, which revealed a small, dense nucleus at the center of the atom. Rutherford's nuclear model proposed that the atom consisted mostly of empty space, with electrons orbiting a central nucleus, much like planets around the sun. Further refinements to the atomic model were made by Niels Bohr, who incorporated quantum theory to explain the stability of Rutherford’s atomic structure. Bohr introduced the concept of quantized orbital shells for electrons, which could explain the emission and absorption spectra of atoms. His model was crucial in advancing the field of quantum mechanics, providing a theoretical basis for the electronic structure of atoms \cite{bohr1934atomic}.

The quantum story began in earnest with Max Planck's groundbreaking proposal in 1900 that energy is quantized, introducing the concept of the quantum to solve the black-body radiation problem. This revolutionary idea was soon followed by Albert Einstein's 1905 explanation of the photoelectric effect, which further emphasized the particle-like properties of light, quantified as photons. These developments challenged the prevailing wave theories of light and prompted a rethinking of the fundamental principles of physical reality. The formal structure of quantum mechanics began to take shape in the 1920s with the contributions of luminaries like Niels Bohr, Werner Heisenberg, and Erwin Schrödinger. Bohr introduced the quantum model of the hydrogen atom in 1913, which described electrons orbiting the nucleus in quantized energy levels, offering a quantum explanation for the atomic stability and spectral lines. Heisenberg's matrix mechanics (1925) and Schrödinger's wave mechanics (1926) provided two mathematically distinct but theoretically equivalent formulations of quantum mechanics, each deepening the understanding of atomic and subatomic processes \cite{heisenberg1973development}.

The new quantum theory rapidly evolved, incorporating the statistical interpretation of the wave function introduced by Max Born, and the uncertainty principle articulated by Heisenberg, which posits fundamental limits to the precision with which certain pairs of physical properties, like position and momentum, can be simultaneously known. These principles underscored the inherent probabilistic nature of quantum measurement outcomes, marking a departure from deterministic classical physics. Throughout the mid-20th century, quantum mechanics was further refined and expanded through the development of quantum field theory, which reconciled quantum mechanics with special relativity and provided a framework for describing particle interactions via fields. This era saw the formulation of theories like quantum electrodynamics by Richard Feynman, Julian Schwinger, and Sin-Itiro Tomonaga, which precisely described the interactions of electrons and photons, and led to predictions of extraordinary accuracy.

\section{Schrödinger Equation}

The Schrödinger equation is a fundamental equation in quantum mechanics that governs the wave function of a quantum-mechanical system. It was postulated by Erwin Schrödinger in 1925 and published in 1926, providing a significant foundation for modern quantum theory. The equation is named after Schrödinger, who received the Nobel Prize in Physics in 1933 for his work. Conceptually, the Schrödinger equation serves as the quantum counterpart to Newton's second law in classical mechanics. While Newton's second law predicts the trajectory of a classical system given initial conditions, the Schrödinger equation predicts the evolution of the wave function over time, encapsulating the quantum state of a system. The equation was inspired by Louis de Broglie's hypothesis that all matter exhibits wave-like properties, leading to the prediction of atomic bound states that matched experimental data.

The general form of the time-dependent Schrödinger equation is given by:

\begin{equation}
i\hbar \frac{\partial}{\partial t} \Psi(\mathbf{r}, t) = \hat{H} \Psi(\mathbf{r}, t)
\end{equation}

Here, \( \Psi(\mathbf{r}, t) \) is the wave function, \( \hat{H} \) is the Hamiltonian operator representing the total energy of the system, \( \hbar \) is the reduced Planck constant, and \( i \) is the imaginary unit \cite{berezin2012schrodinger}.

For a single non-relativistic particle in one dimension, the Schrödinger equation can be written as:

\begin{equation}
i\hbar \frac{\partial}{\partial t} \Psi(x, t) = \left[ -\frac{\hbar^2}{2m} \frac{\partial^2}{\partial x^2} + V(x, t) \right] \Psi(x, t)
\end{equation}

In this equation, \( m \) is the mass of the particle, \( V(x, t) \) is the potential energy, and \( \frac{\partial^2}{\partial x^2} \) represents the second derivative with respect to position \cite{karam2020schrodinger}.

In cases where the Hamiltonian does not explicitly depend on time, the wave function can be separated into spatial and temporal components, leading to the time-independent Schrödinger equation:

\begin{equation}
\hat{H} \psi(\mathbf{r}) = E \psi(\mathbf{r})
\end{equation}

Here, \( \psi(\mathbf{r}) \) is the spatial part of the wave function, and \( E \) represents the energy eigenvalue of the system.

The Schrödinger equation was a significant milestone in the development of quantum mechanics, offering a new way to understand the behavior of microscopic systems. Although Schrödinger initially attempted to interpret the wave function \( \Psi \) as representing charge density, Max Born later provided the correct interpretation: the modulus squared of the wave function \( |\Psi|^2 \) represents the probability density of finding a particle in a given state \cite{bernstein2005max}. This interpretation remains central to quantum mechanics today. The Schrödinger equation is non-relativistic, as it contains a first derivative in time and a second derivative in space, treating space and time asymmetrically. This limitation is addressed in relativistic quantum mechanics by the Klein-Gordon and Dirac equations, which incorporate special relativity \cite{barukvcic2013relativistic}. The Dirac equation, in particular, reduces to the Schrödinger equation in the non-relativistic limit.

\section{Quantum Field Theory}

Quantum Field Theory (QFT) represents a fundamental framework in theoretical physics, combining elements of classical field theory, quantum mechanics, and special relativity. This powerful theory is crucial for our understanding of particle physics and has applications in condensed matter physics and other areas of physics. Quantum fields are seen as the fundamental building blocks of the universe, with particles acting as excitations of these fields \cite{ryder1996quantum}. In QFT, particles such as electrons and photons are not described as distinct points but rather as excited states of underlying fields that permeate all of space \cite{peskin2018introduction}. For example, the electron is an excitation of the electron field, while the photon is an excitation of the electromagnetic field. This conceptual shift from particles to fields helps address phenomena that are inexplicable by classical physics, such as the interactions between light and matter and the creation and annihilation of particles \cite{tait1972quantum}.

The equations of motion in QFT are derived from a Lagrangian that includes terms for each field and interactions among fields \cite{zwanziger1971local}. These interactions are depicted graphically by Feynman diagrams, which provide a visual and mathematical way to calculate the probabilities of various physical processes. The interaction terms in the Lagrangian typically involve products of field operators, and the dynamics of these fields are governed by the principles of quantum mechanics and special relativity. Feynman diagrams are a powerful and intuitive tool used in quantum field theory to represent the complex interactions between subatomic particles \cite{argeri2007feynman}. A typical Feynman diagram includes external lines, internal lines, and vertices. External lines represent the incoming and outgoing particles in a process, such as an electron entering a reaction or a photon being emitted. Internal lines depict virtual particles—temporary, intermediate particles that exist only fleetingly during the interaction and connect different vertices within the diagram. Vertices, the points where lines meet, represent the interaction points where particles either exchange other particles or transform into different ones. For example, in quantum electrodynamics (QED), the simplest Feynman diagram might show an electron emitting or absorbing a photon, with the interaction represented by a vertex where an electron line meets a wavy line representing the photon.

Feynman diagrams are not merely illustrative; they are a fundamental tool for calculating the probabilities of various particle interactions \cite{meynell2008feynman}. Each line and vertex in a diagram corresponds to a specific mathematical term in the overall equation that describes the process. The translation of a Feynman diagram into a mathematical expression follows specific ``Feynman rules," which vary depending on the particular quantum field theory being used \cite{christensen2009feynrules}. The elegance of Feynman diagrams lies in their ability to encapsulate the entirety of a particle interaction in a single, often simple, visual representation. Even complex interactions involving multiple particles and forces can be broken down into simpler diagrams, which can then be calculated using these rules. The total probability amplitude for a given process is determined by summing all possible diagrams that represent that process \cite{graham1977path}.

\section{Path Integral Formulation}

The path integral formulation of quantum mechanics is a powerful framework that generalizes the classical principle of stationary action to include quantum phenomena. Unlike the classical description, where a system follows a single, unique trajectory, the path integral approach considers an infinite number of possible paths that a system can take between two points in spacetime. Each path contributes to the quantum amplitude, with a phase factor determined by the action along that path. This method was developed by Richard Feynman in 1948 and has since become a cornerstone of modern theoretical physics.

In this formulation, the quantum amplitude for a particle to move from point A at time \( t_A \) to point B at time \( t_B \) is given by summing over all possible paths connecting these points. The contribution of each path is weighted by \( e^{iS/\hbar} \), where \( S \) is the action calculated along that path. Mathematically, this can be expressed as:

\begin{equation}
\langle B | A \rangle = \int \mathcal{D}[x(t)] e^{iS[x(t)]/\hbar}
\end{equation}

Here, \( \mathcal{D}[x(t)] \) represents the measure over all possible paths \( x(t) \), and \( S[x(t)] \) is the classical action, which is a functional of the path \cite{feynman1965path}.

In the context of quantum mechanics, the path integral formulation offers deep insights into phenomena like quantum tunneling and the behavior of particles in potential fields. For instance, the quantum tunneling rate can be derived by evaluating the path integral for paths that traverse potential barriers, yielding results consistent with experimental observations \cite{tanizaki2014real}. However, the path integral formulation also comes with challenges. The functional integral over an infinite number of paths is often difficult to evaluate directly, and various approximations, such as the stationary phase approximation, are employed. Moreover, ensuring the unitarity of the S-matrix (which corresponds to the conservation of probability) is less transparent in this formulation and requires careful handling \cite{haba1994stochastic}.

\section{Uncertainty principle}
The Uncertainty Principle, introduced by Werner Heisenberg in 1927, is a fundamental theory in quantum mechanics. It asserts a fundamental limit to the precision with which pairs of physical properties, such as position ($x$) and momentum ($p$), can be simultaneously known. This principle has profound implications across quantum physics, highlighting the inherent limitations of our measurement capabilities at the quantum scale.

The principle is best known in one of its mathematical forms, which relates the standard deviation of position $\sigma_x$ and the standard deviation of momentum $\sigma_p$:
\begin{equation}
\sigma_x \sigma_p \geq \frac{\hbar}{2}
\end{equation}
where $\hbar = \frac{h}{2\pi}$ is the reduced Planck constant.

This fundamental limit arises because every particle's position and momentum are linked by their wave properties. In quantum mechanics, every particle is also a wave, and its position and momentum are described by a wave function $\psi(x)$. The precision in determining the position and momentum of a particle is governed by the spread of its wave function and its Fourier transform, respectively. A more precisely defined position leads to a broader spread in momentum and vice versa. This is formalized by the Fourier transform properties of the wave functions \cite{slepian1961prolate}:
\begin{equation}
\psi(x) = \int e^{-ipx/\hbar} \phi(p) \, dp
\end{equation}
and
\begin{equation}
\phi(p) = \int e^{ipx/\hbar} \psi(x) \, dx.
\end{equation}

The Uncertainty Principle is not merely a statement about the observational limits imposed by our current technology; rather, it is a fundamental property of the universe. It applies not only to position and momentum but to other pairs of physical properties as well, such as energy and time, for which a similar uncertainty relation can be derived.

The energy-time uncertainty relation is similarly interpreted:
\begin{equation}
\Delta E \Delta t \geq \frac{\hbar}{2}
\end{equation}
This relation implies that the measurement of energy in a quantum system over a finite time interval $\Delta t$ cannot be more precise than $\hbar/2$ divided by that interval.

\section{Quantum Entanglement}

Quantum entanglement is a fundamental and counterintuitive phenomenon in quantum mechanics, wherein the quantum states of two or more particles become intertwined such that the state of each particle cannot be described independently of the state of the others, regardless of the distance separating them. This phenomenon illustrates a profound departure from classical physics, where objects are considered to have distinct, independent properties regardless of their interaction history. The concept of quantum entanglement can be traced back to the early 20th century when Albert Einstein, Boris Podolsky, and Nathan Rosen (EPR) presented a thought experiment challenging the completeness of quantum mechanics \cite{bell1964einstein}. The EPR paradox posited that if quantum mechanics were correct, then measuring the state of one particle in an entangled pair instantaneously determines the state of the other, even if the particles are light-years apart \cite{kupczynski2016epr}. Einstein famously referred to this as ``spooky action at a distance," expressing his discomfort with the idea that information could seemingly travel faster than the speed of light, thus violating the principle of locality. Despite Einstein's reservations, subsequent theoretical work and experiments, most notably those conducted by John Bell in the 1960s, demonstrated that the predictions of quantum mechanics—specifically, the strong correlations between entangled particles—could not be explained by any local hidden variable theory \cite{wiseman2014two}. Bell's theorem provided a means to test the differences between quantum mechanics and local realism, leading to a series of experiments that confirmed quantum mechanics' predictions and invalidated local hidden variable theories. The experimental violation of Bell's inequalities has since been demonstrated in numerous ``loophole-free" experiments, reinforcing the non-local nature of quantum entanglement.

At the heart of entanglement lies the principle that the combined quantum state of a system of particles can be described as a superposition of all possible states, where each particle's state is intrinsically linked to the others. This entangled state is such that measuring one particle's property (such as spin, polarization, or position) instantaneously affects the state of the other particle, collapsing the superposition into a definite outcome. This collapse occurs simultaneously for all entangled particles, no matter the distance between them, a feature that defies classical intuitions about space and time. Entanglement has been observed in a variety of systems beyond photons, including electrons, atoms, and even macroscopic objects under carefully controlled conditions.

\section{Spacetime}

Spacetime is a foundational concept in modern physics, unifying the dimensions of space and time into a single, four-dimensional continuum. This concept revolutionized the way we understand the universe, particularly through the theories of special and general relativity. In classical mechanics, space and time were treated as separate entities. Space was a three-dimensional stage where events occurred, while time flowed uniformly and independently of space. The theory of relativity, developed by Albert Einstein, signifies a pivotal transformation in modern physics, encompassing two theories: special relativity and general relativity. Special relativity, introduced in 1905, redefined the concepts of time and space, which were previously considered as independent and absolute. Contrary to these earlier notions, Einstein proposed a spacetime continuum where time and space are interlinked and relative to the observer. Special relativity rests on two fundamental postulates: the laws of physics are invariant across all inertial frames of reference, and the speed of light in a vacuum is a constant, unaffected by the motion of the source or the observer. These postulates introduce phenomena such as time dilation, where time appears to slow down for an object in motion relative to a stationary observer; length contraction, where objects in motion are observed to be shorter along the direction of motion; and mass-energy equivalence, expressed by the equation \(E=mc^2\), asserting that mass and energy are interchangeable \cite{choquet2009general}.

In 1915, Einstein extended these principles through his formulation of general relativity, which includes the effects of gravity on spacetime. Central to general relativity is the principle of equivalence, which posits that the effects of gravity are indistinguishable from those of acceleration. Hence, gravity is not described as a conventional force but as a manifestation of the curvature of spacetime caused by mass and energy. This curvature dictates the trajectories of objects, which move along paths known as geodesics \cite{hawking2010general}. General relativity's predictions have been substantiated through various experimental and observational confirmations. These include gravitational lensing, where light bends around massive objects; the anomalous precession of Mercury's orbit, unaccounted for by Newtonian mechanics; and the detection of gravitational waves—ripples in spacetime generated by cataclysmic events such as mergers of black holes, a phenomenon confirmed by the LIGO observatory \cite{berti2015testing,martynov2016sensitivity}. Spacetime is a dynamic entity in general relativity. Massive objects like stars and black holes warp the spacetime around them, creating gravitational fields. To visualize these effects, physicists use the concept of spacetime. In this four-dimensional framework, events are specified by three spatial coordinates (x, y, z) and one time coordinate (t). A spacetime diagram, also known as a Minkowski diagram, helps illustrate how different observers perceive events differently based on their relative velocities \cite{naber2012geometry}. The unification of space and time into spacetime shows that the separation of these two entities is not absolute but depends on the observer’s state of motion.

\section{Delayed-Choice Quantum Eraser}

The delayed-choice quantum eraser experiment is an advanced and intriguing variation of the quantum eraser experiment, which itself is a derivative of the famous double-slit experiment in quantum mechanics. First performed by Yoon-Ho Kim, R. Yu, S. P. Kulik, Y. H. Shih, and Marlan O. Scully in 1999, this experiment combines the principles of quantum entanglement with the concept of a delayed choice, originally proposed by physicist John Archibald Wheeler \cite{yoon2000delayed}. The delayed-choice quantum eraser experiment explores the perplexing consequences of quantum mechanics, specifically addressing the nature of wave-particle duality and the role of the observer in determining quantum states.

\begin{figure}[ht]
\centering
\includegraphics[width=\textwidth]{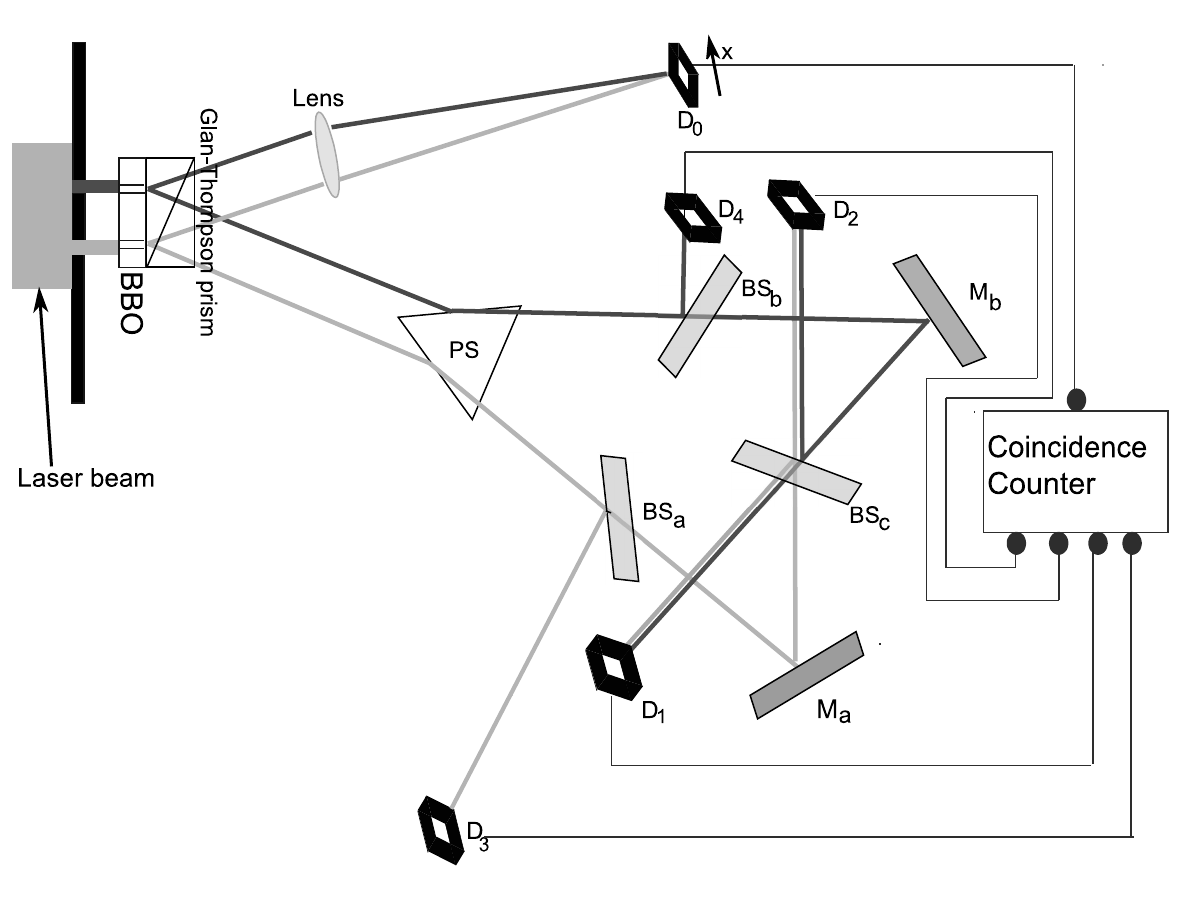}
\caption{Delayed-Choice Quantum Eraser experiment setup \cite{moran_own_work}}
\label{fig:eraser}
\end{figure}

The delayed-choice quantum eraser experiment is an extension of the double-slit experiment, where a beam of light, usually from a laser, is directed towards a barrier with two parallel slits. When light passes through these slits, it can produce an interference pattern on a detection screen, indicative of wave-like behavior. However, if a detector is placed at the slits to observe which slit the photon passes through, the interference pattern disappears, and the light behaves like a particle. This phenomenon highlights the principle of complementarity in quantum mechanics: a quantum system can display either wave-like or particle-like properties depending on the experimental setup. In the delayed-choice quantum eraser, the setup is more complex. After photons pass through the slits, they are subjected to spontaneous parametric down-conversion (SPDC), a process that generates pairs of entangled photons. One of these photons, called the ``signal photon," proceeds towards a primary detector (D0), while the other, known as the ``idler photon," is sent towards a series of detectors (D1, D2, D3, and D4) positioned at varying distances and with different configurations of beam splitters and mirrors. The key aspect of the experiment is that the idler photons, which could provide ``which-path" information (i.e., knowledge about which slit the photon passed through), are detected after the signal photon has already been recorded at D0. This introduces a ``delayed choice" element: the decision to preserve or erase the which-path information is made after the signal photon has been detected.

The experiment yields results that challenge classical intuitions about time and causality. When the idler photons are detected at D3 or D4, which provide which-path information, the signal photons at D0 do not form an interference pattern; they behave as if they had traveled through one specific slit, exhibiting particle-like behavior. Conversely, when the idler photons are detected at D1 or D2, where which-path information is effectively ``erased," the signal photons at D0 display an interference pattern, indicative of wave-like behavior. What makes this result remarkable is that the interference pattern (or lack thereof) at D0 seems to be determined by a choice made after the signal photon has already been detected. This appears to imply that the measurement made on the idler photon retroactively influences the outcome of the signal photon, even though the events are separated in time. The delayed-choice quantum eraser experiment suggests that quantum systems do not have definite properties (such as being a wave or a particle) until they are observed, and that the nature of these properties can be influenced by measurements made after the fact.

\section{Particle Fallacy}

Before delving into the intricacies of the Delayed-Choice Quantum Eraser experiment, it is crucial to address and dispel some widespread misconceptions, particularly the ``particle fallacy." This fallacy arises from the classical intuition that electrons and other subatomic entities are discrete, localized particles moving along defined trajectories. However, according to the Schrödinger model of quantum mechanics, this view is fundamentally flawed. In the Schrödinger model, the electron is not a particle in the traditional sense but rather a quantum entity best described by a wave function—a probability density function spread over its eigenstate. This wave function represents the electron's state, encapsulating all possible outcomes of a measurement, such as position or momentum, as probabilities rather than certainties. The notion of the electron as a particle with a definite position and path is an approximation that only emerges under specific conditions, such as when a measurement collapses the wave function to a particular eigenstate. Outside of these measurements, the electron exists in a superposition of states with its properties being fundamentally indeterminate. 

The wave function in quantum mechanics is often described as a mathematical tool used to calculate probabilities of finding a particle in a particular state or position. However, this description overlooks the deeper physical significance of the wave function, particularly in the context of quantum field theory. In QFT, particles are understood as excitations of underlying quantum fields that permeate all of space. The wave function, which describes the quantum state of a particle, is not merely an abstract mathematical construct; rather, it is intimately tied to the physical reality of these fields. The wave function represents the state of the quantum field itself, and its evolution is governed by the Schrödinger equation or its relativistic equivalents, such as the Dirac or Klein-Gordon equations. The physical manifestation of the wave function becomes particularly evident in phenomena like interference and diffraction, where the wave-like nature of particles—such as electrons or photons—is observable. 

It is essential to clarify that Feynman diagrams should not be interpreted as depicting individual paths that a particle might take, as this perspective can be misleading. The traditional view, where each diagram represents a specific trajectory or path taken by a particle, falls into the trap of classical thinking, which does not apply to the quantum realm. In quantum mechanics, there are no particles following definite paths through space and time. Instead, we must understand the concept of the wave function—a quantized probability amplitude that is distributed across all possible paths and configurations simultaneously. The wave function encapsulates all possible outcomes and states that a quantum system can inhabit, representing a superposition of these possibilities rather than a single, defined trajectory. When we use Feynman diagrams, they serve as a powerful computational tool to represent the sum of all possible interactions and processes that contribute to the overall probability amplitude of a quantum event. Each diagram is a graphical representation of a term in the perturbative expansion of the wave function, contributing to the total amplitude in a manner that reflects the quantum superposition of all possible states and interactions. The diagrams illustrate how these interactions contribute to the overall wave function, not as discrete alternatives but as components of a unified, non-classical quantum reality. In essence, Feynman diagrams do not depict individual paths but rather the collective sum of all potential quantum interactions, where the wave function is distributed across all possibilities. This understanding aligns with the core principles of quantum mechanics, where the concept of a particle following a single path is replaced by the more accurate description of a wave function that encompasses all possibilities at once.

\section{The Confinement of the Wave Function}

In quantum mechanics, the act of measurement fundamentally alters the system being observed, particularly the wave function that describes the system's state. Before measurement, the wave function is a superposition of all possible states, spread out across all potential paths and configurations. This wave function is not localized or confined; rather, it represents a probabilistic distribution over all possible outcomes that the system can achieve. However, when we attempt to measure the wave function, we are forced to interact with it using some form of measurement device or particle. This interaction is not passive; it imposes constraints on the wave function, effectively confining it to a specific region of space or a particular eigenstate. The act of measurement restricts the wave function to those paths that directly interact with the measurement apparatus, thereby collapsing the wave function from a superposition of states into a single, observable outcome. This confinement of the wave function is analogous to constraining a crowd of people. Imagine a crowd dispersing freely in an open space, representing the unmeasured wave function. As soon as you introduce a gate through which they must pass, you limit the directions they can take. Similarly, when a measurement device interacts with a quantum system, it acts like a gate, restricting the system's wave function to those paths and states that can pass through this ``gate."

\begin{figure}[ht]
\centering
\includegraphics[width=\textwidth]{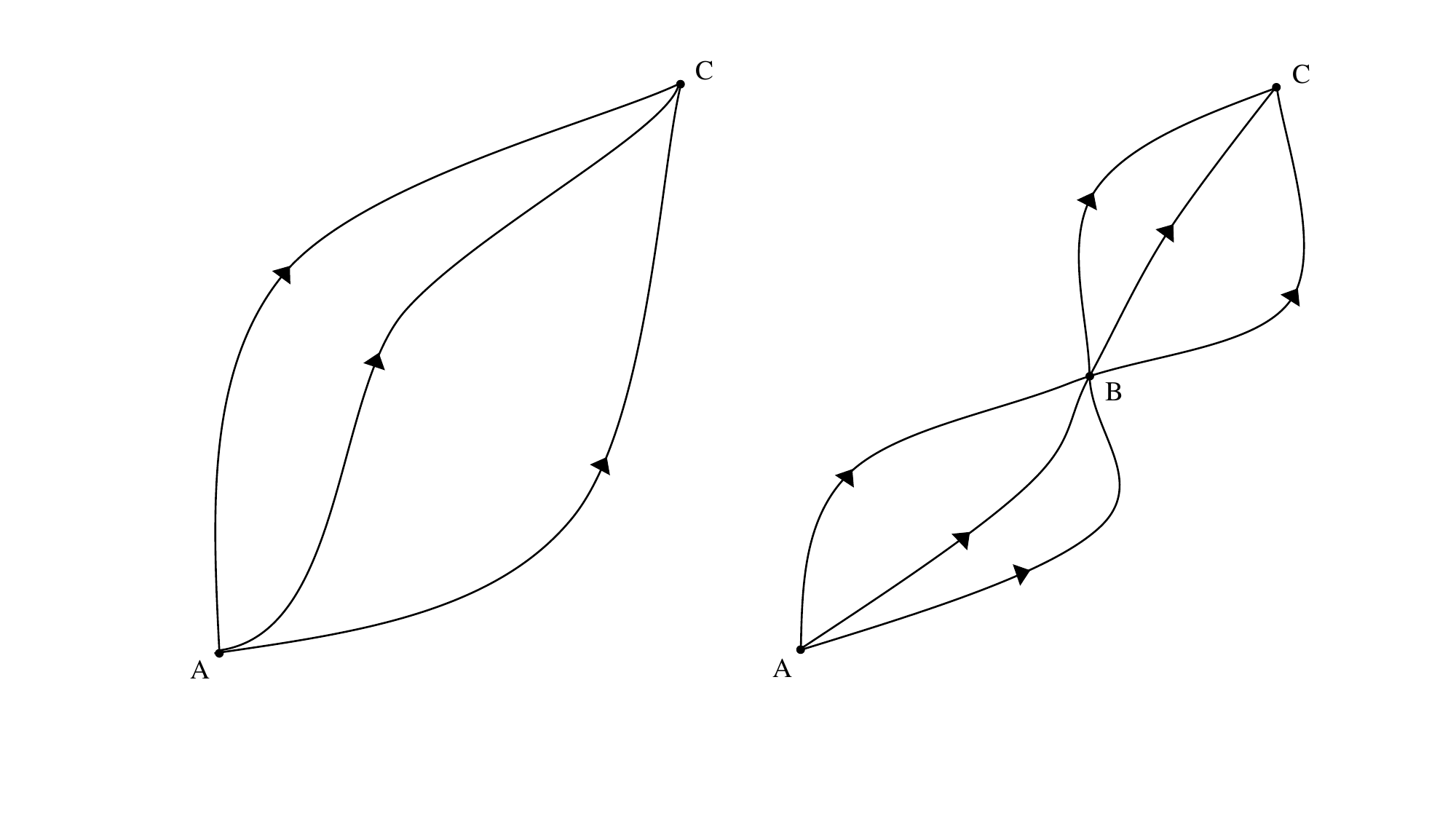}
\caption{The introduction of point B constrains all possible paths between A and C by requiring that they must pass through B, thereby reducing the set of valid paths to those that include both segments A-B and B-C.}
\label{fig:paths}
\end{figure}

In terms of Feynman Paths, which represent the sum of all possible quantum interactions, the act of measurement eliminates many of these possible paths. Only the paths consistent with the measurement constraints remain part of the solution space. The other trajectories—the ones that would have been possible in the absence of measurement—are no longer relevant to the observed outcome. In this sense, measurement not only changes the state of the system but also reduces the complexity of the quantum interactions we must consider. By collapsing the wave function, measurement simplifies the system to a single trajectory, effectively discarding the alternative paths and possibilities that were initially part of the quantum superposition. This inherent limitation underscores a key principle of quantum mechanics: the very act of measuring a system changes it. The wave function, which initially encompasses all possible states, is forced to conform to the constraints imposed by the measurement. This process reveals one of the fundamental limitations in our ability to fully understand a quantum system: we can only observe the state that remains after measurement, not the full array of possibilities that existed beforehand.

Unlike classical systems, where transitions occur gradually and continuously from one state to another, quantum systems exhibit what can be described as instantaneous transitions between eigenstate. There are no intermediate or ``in-between" states during these transitions; the wave function moves directly from one eigenstate to another without passing through any transitional phases. This instantaneous transition is deeply tied to the probabilistic nature of quantum mechanics. The wave function of a quantum system represents a superposition of all possible states, each associated with a certain probability amplitude. When a measurement is made, the wave function collapses instantaneously into one of these possible states. The likelihood of the wave being collapsed in any particular eigenstate is directly proportional to the probability density function derived from the wave function. This density function tells us where the system, such as an electron, is most likely to be found upon measurement, but before the measurement, the electron is not in any single place—its existence is spread out across all potential locations.

Consider the double-slit experiment as an example. When an electron passes through the slits, its wave function spreads out, encompassing all possible paths and positions it could occupy. However, when it is detected on the screen behind the slits, the wave function collapses/absorbs instantly, and the electron is found in a specific location on the screen. This location corresponds to one of the many possible outcomes dictated by the wave function's probability distribution. Keep in mind the the detection point is most probably an atom that can host the wave function in one of its orbitals. Electron was never a particle here, it just transitioned from a wave function in space to the wave function in orbital. Moreover, this process of wave function collapse is governed entirely by the probability density function at the specific location of detection. If the wave function interacts with a particle or atom at that location, the interaction is instantaneous, and the system is immediately confined to a new eigenstate. This highlights the fact that quantum mechanics does not allow for smooth, continuous evolution between states during measurement; rather, it involves abrupt, discrete changes governed by the probabilistic nature of the wave function.

It's crucial to understand that when discussing quantum mechanics, especially in the context of experiments like the double-slit experiment, we must discard any anthropomorphic notions about particles ``choosing" paths or making decisions. Particles do not possess consciousness or any form of agency that would allow them to ``decide" which slit to pass through or how to behave when observed. The behavior of particles is purely a result of the mathematical structure of quantum mechanics, which describes the evolution and measurement of the wave function. In the double-slit experiment, the particle’s wave function initially spreads out to encompass all possible paths through both slits. When no measurement device is present to detect which slit the particle goes through, the wave function continues to evolve as a superposition of all these possibilities, leading to an interference pattern on the detection screen—a hallmark of the wave-like behavior of quantum particles. However, when a measurement device is introduced—such as a detector placed at one of the slits—the situation changes dramatically. The act of measurement forces the wave function to collapse into a state that corresponds to the particle having passed through one slit or the other. Remember, half of the wave cannot collapse. The wave is quantized and belongs to a single particle, so the collapse is an all-or-none phenomenon. This is why placing a measuring device by the slit causes the wave to pass through one slit or the other. This collapse is not a choice made by the particle but a direct consequence of the interaction between the wave function and the measurement device. The wave function is forced into an intermediate eigenstate, determined by the constraints of the measurement apparatus. Importantly, once the gate (or measurement device) is removed, the wave function is no longer constrained by the need to interact with the measurement apparatus. The wave function then resumes its original, uncollapsed form, spreading out once more to encompass all possible paths through both slits.

\section{Time Uncertainty}

In this section, we aim to reinterpret the Uncertainty Principle by exploring its implications within the spacetime domain. Starting from the conventional expression relating the uncertainties in position and momentum, by considering momentum as the product of mass and velocity, and assuming that the mass of the particle remains constant, we will show that the uncertainty in momentum translates directly into uncertainty in velocity. Since velocity can be understood as the uncertainty in position over time, we will then express the Uncertainty Principle in terms of the uncertainties in both space and time. This reformulation underscores the interconnected nature of space and time at the quantum level, offering a more comprehensive view of the constraints imposed by quantum mechanics. 

To explore this concept further, let's consider the momentum of a particle. Momentum (\(p\)) is defined as the product of mass (\(m\)) and velocity (\(v\)):

\begin{equation}
p = mv
\end{equation}

If we assume that the mass of the particle is constant and well-defined (i.e., there is no uncertainty associated with the mass), any uncertainty in momentum (\(\sigma_p\)) must therefore arise from the uncertainty in the velocity (\(\sigma_v\)) of the particle:

\begin{equation}
\sigma_p = m\sigma_v
\end{equation}

Given that velocity is the rate of change of position with respect to time, the uncertainty in velocity can be understood as the uncertainty in position (\(\sigma_x\)) over a given time interval (\(\sigma_t\)):

\begin{equation}
v = \frac{\Delta x}{\Delta t} \quad \text{or} \quad \sigma_v = \frac{\sigma_x}{\sigma_t}
\end{equation}

Substituting this into the expression for \(\sigma_p\), we get:

\begin{equation}
\sigma_p = m \frac{\sigma_x}{\sigma_t}
\end{equation}

Now, considering the Uncertainty Principle in terms of space and time, we can rewrite the principle by substituting \(\sigma_p\) from the above equation:

\begin{equation}
\sigma_x \left(m \frac{\sigma_x}{\sigma_t}\right) \geq \frac{\hbar}{2}
\end{equation}

Simplifying this expression, we obtain:

\begin{equation}
\frac{\sigma_x^2}{\sigma_t} \geq \frac{\hbar}{2m}
\end{equation}

This equation illustrates that the uncertainty in position (\(\sigma_x\)) is directly related to the uncertainty in the time interval (\(\sigma_t\)) over which the position is measured, given a constant mass. In this sense, the Uncertainty Principle can be interpreted as an uncertainty relation in spacetime, where the precision with which we can measure the position of a particle is fundamentally linked to the time interval during which the measurement occurs. To get around the inaccuracies of this derivation, we can assume the Uncertainty Principal holds a general form of:

\begin{equation}
\sigma(x,t) \geq c
\end{equation}

which \(\sigma(x,t)\) signifies the uncertainty in spacetime, while $c$ is a constant.

Another crucial aspect to consider is the effect of time dilation when a particle is traveling close to the speed of light. The reference time for such a particle is not the same as that for a stationary observer. Due to relativistic effects, time for the particle slows down relative to the observer's frame of reference. In the quantum world, the concept of ``present time" as we understand it in our everyday experience does not have the same meaning. The notion of a universal ``now," where events are occurring simultaneously across the universe, is an artifact of our classical understanding of time, deeply rooted in our human perception. However, this concept falls apart when we delve into the quantum realm.

In quantum world, each wave function exists within its own spacetime domain, governed by the principles of quantum fields. The wave function is not confined to a single moment or location; instead, it spans across the entirety of spacetime, encompassing all possible states and interactions. This means that what we refer to as ``the present" is not a definitive, universally shared moment but rather a construct that emerges from our macroscopic perspective. For a wave function, time and space are intertwined in such a way that there is no single ``present" but rather a continuous distribution of possibilities across the entire spacetime. Each wave function exists within this spacetime, with its evolution governed by the underlying quantum field. The outcomes we observe, which we might describe as occurring in ``the present," are actually the result of interactions between wave functions across different points in spacetime.

\section{Wave Function in Spacetime}

When a particle such as an electron or photon passes through the two slits, its wave function splits into two distinct wave components, each corresponding to one of the slits. These wave components then propagate and overlap on the other side of the slits. The overlap of these wave components results in the phenomenon of interference, which can be either constructive (where the wave components reinforce each other) or destructive (where they cancel each other out). This interference of the wave function creates a pattern of alternating high and low probabilities on the detection screen placed behind the slits.

\begin{figure}[ht]
\centering
\includegraphics[width=\textwidth]{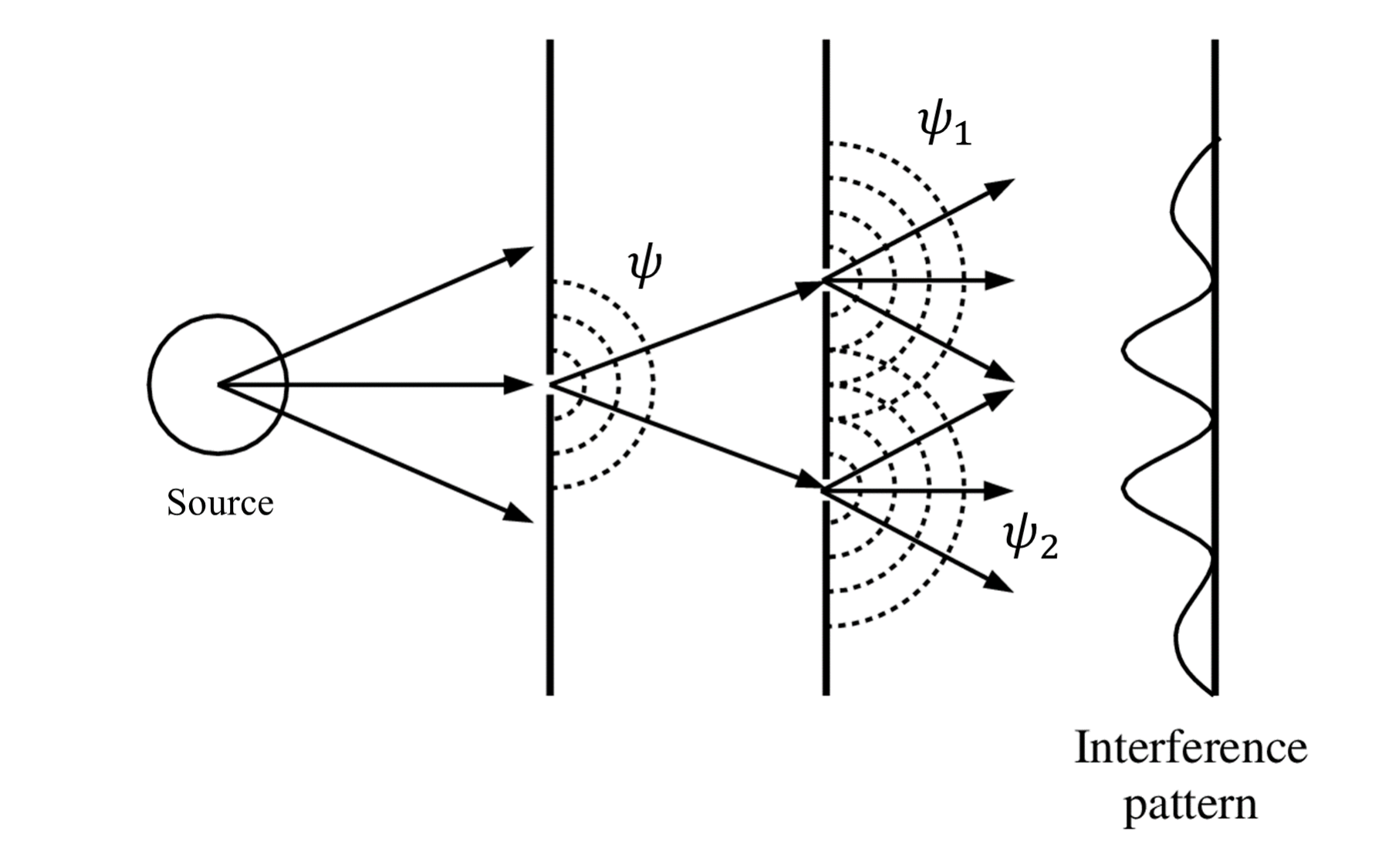}
\caption{Interference of single particle wave function (background image is captured from \cite{davis_interference}).}
\label{fig:ip}
\end{figure}

When a particle is detected on the screen, the wave function does not manifest in a continuous distribution across the screen but rather collapses to a single point. The position where the particle is detected is determined stochastically, meaning that it is random but follows the probability distribution given by the square of the wave function's amplitude. This process leads to the formation of an interference pattern over time, as more and more particles are detected at the screen. The pattern consists of regions with high and low densities of detected particles, corresponding to the constructive and destructive interference of the wave function, respectively. This quantum mechanical interpretation highlights that the interference pattern observed in the double-slit experiment emerges from the wave-like nature of the particle's wave function and the probabilistic nature of quantum mechanics. The particle itself does not interfere with itself in a classical sense; rather, the wave function that describes the particle's quantum state exhibits interference, and the detection events occur at positions that reflect this interference pattern.

In the context of the Delayed-choice quantum eraser experiment, it's essential to understand that the wave function does not merely spread across space but across spacetime as a whole. The wave function is a mathematical representation of the quantum state of a system, and this state encompasses not just the spatial distribution of possibilities but also their temporal evolution. To fully capture the behavior of quantum systems, one must consider the wave function as a solution to the Schrödinger equation that is valid across both space and time. This means that the wave function evolves over time, and its admissible solutions are governed by the constraints of both spatial boundaries and temporal ones. The wave function must satisfy these conditions to provide a coherent description of the quantum system.

It is constructive to think of entangled particles as entangled wave functions within a quantum field, which ensures that the energy, momentum, and spin of the joint wave remain consistent, preserving underlying field and spacetime symmetries, provided they do not interact with other wave functions. This approach emphasizes the wave-like nature of quantum entities and how their properties are governed by the overall quantum field, maintaining entanglement as long as external interactions are minimized. The particles involved are not just entangled through space, but through both time and space. This entanglement arises because the particles are manifestations of the same underlying wave function, which is a solution to the quantum field governing the system. The quantum field, which dictates the behavior of particles and their interactions, ensures that the wave function remains consistent and admissible across the entirety of spacetime. Entanglement in this context means that the properties of one particle are intrinsically linked to those of another, regardless of the distance separating them. The wave function's evolution through spacetime allows for the intriguing possibility that actions taken at a later time can seemingly affect outcomes observed at an earlier time. This is not because the wave function is ``deciding" anything but because it exists as a continuous, spacetime-spanning entity. The entire history of the wave function, including the interaction with measurement devices, must conform to the allowable solutions of quantum mechanics. Therefore, the wave function's spread through spacetime incorporates all possible paths and interactions, constrained by the conditions imposed by the experiment.

When we measure and confine a wave function, we are not only determining the state of the particle associated with that wave function, but we are also influencing its entangled counterpart across spacetime. This confinement, imposed by the act of measurement, forces the wave function to ``collapse" into a specific eigenstate within the bounds of the measurement interaction. Consequently, this collapse is not an isolated event; it is intricately connected to the state of the entangled particle, regardless of the distance or time separating them. In the realm of quantum mechanics, entangled particles share a single, unified wave function that extends across spacetime. This means that the wave function of one particle cannot be described independently of its entangled partner. When we interact with one particle and confine its wave function to a particular state, this interaction immediately imposes constraints on the wave function of the other particle. As a result, other potential paths that the wave function might have taken—paths that do not satisfy the constraints imposed by the measurement—are no longer admissible within the spacetime framework. These non-admissible paths are effectively eliminated from the wave function, both in space and time.

This process might give the illusion that particles are somehow communicating with their counterparts in the future, as if they possess consciousness or agency to influence one another across time, or if our consciousness affects the physical reality. However, it is essential to understand that particles do not have any form of consciousness or ability to make choices. The phenomenon we observe is purely a consequence of the constraints enforced by the quantum field within spacetime. The misconceptions form when we want to probe 4D space with our limited 3D senses/devices. When a wave function is measured and confined, it is not because the particle is ``aware" of its entangled partner or communicates with it. Rather, the constraints imposed by the measurement simply dictate the only allowable solution in spacetime for the entire quantum system. The entangled partner's wave function is then automatically adjusted to maintain consistency with the laws of quantum mechanics. This adjustment is not a result of communication between particles but a reflection of the fundamental nature of quantum mechanics, where the wave function must satisfy all admissible solutions in both space and time. The elimination of non-admissible paths is a consequence of the quantum field ensuring that only those trajectories that fit within the enforced constraints remain.

\section{Conclusion}
The limitations inherent in our measurement devices, combined with the deeply ingrained dogmas of classical physics, create significant barriers to perceiving the true nature of quantum reality. Our instruments, no matter how advanced, are ultimately grounded in the classical framework—they are designed to measure discrete, particle-like events in a well-defined spacetime. However, quantum mechanics operates on principles that defy this classical outlook. The wave function, which encapsulates the quantum state of a system, is not a simple, localized entity but rather a spread-out, probabilistic distribution that exists across spacetime. Our measurements, which are fundamentally classical interactions, inevitably collapse this wave function, giving us only a partial and often misleading view of the quantum world. Moreover, our classical physics background has trained us to think in terms of distinct particles moving along well-defined paths in space and time. This perspective, while effective for macroscopic phenomena, is inadequate for understanding quantum mechanics, where particles do not have a definite position or momentum until they are observed. The classical notion of determinism is replaced by probabilities, and the clear-cut distinction between past, present, and future is blurred in the quantum realm.

\bibliographystyle{unsrt}
\bibliography{references}

\end{document}